\documentclass[a4paper,floatfix,amsmath,amssymb,pre,showpacs,twocolumn,superscriptaddress,nofootinbib]{revtex4-1}
\usepackage{amsmath}
\usepackage{graphicx}

\usepackage{subfig}
\usepackage{hyperref}
\hypersetup{
pdfstartview=FitV,
colorlinks=true, 
linkcolor=blue, 
citecolor=blue, 
filecolor=black, 
urlcolor=blue,
pdftitle={Surface morphology of a modified ballistic deposition model}}
\usepackage{dcolumn}
\usepackage{bm}
\usepackage{color}

\newcommand{\newc}{\newcommand}
\newc{\mbf}{\mathbf}
\newc{\boma}{\boldmath}
\newc{\phihat}{\mbox{\boldmath{$\hat{\phi}$}}}
\newc{\thetahat}{\mbox{\boldmath{$\hat{\theta}$}}}
\newc{\beq}{\begin{equation}}
\newc{\eeq}{\end{equation}}
\newc{\beqar}{\begin{eqnarray}}
\newc{\eeqar}{\end{eqnarray}}
\newc{\beqa}{\begin{eqnarray*}}
\newc{\eeqa}{\end{eqnarray*}}
\newc{\bd}{\begin{displaymath}}
\newc{\ed}{\end{displaymath}}

\begin{document}

\title{Surface morphology of a modified ballistic deposition model}

\author{Kasturi Banerjee} 
\email{kasturi01@gmail.com}
\affiliation{B.~N.~S.~Girls High School, Howrah 711 303, India}
\author{J. Shamanna}
\email{jlsphy@caluniv.ac.in}
\affiliation{Department of Physics, University of Calcutta, 
Kolkata 700 009, India}
\author{Subhankar Ray}
\email{sray.ju@gmail.com}
\affiliation{Department of Physics, Jadavpur University, 
Calcutta 700 032, India}

\pacs{81.05.Rm, 68..35.-p, 68.35.Ja}
\date{\today}

\begin{abstract}
\noindent The surface and bulk properties of a modified ballistic deposition model are investigated. The deposition rule interpolates between nearest and next-nearest neighbor ballistic deposition and the random deposition models. The stickiness of the depositing particle is controlled by a parameter and the type of inter-particle force. Two such forces are considered - Coulomb and van der Waals type. The interface width shows three distinct growth regions before eventual saturation. The rate of growth depends more strongly on the stickiness parameter than on the type of inter-particle force. However, the porosity of the deposits is strongly influenced by the inter-particle force. 
\end{abstract}

\maketitle

\section{Introduction}
\label{intro}
Surface and structural properties of deposition aggregates are of 
multidisciplinary interest. Deposition structures
occur in various physical, chemical and biological systems 
and processes. The surface and bulk properties of deposition
aggregates are closely related to a wide variety of 
equivalent problems, such as, fluid flow, adsorption and diffusion in 
porous structures, directed polymers in random 
media and propagation of flame fronts \cite{family1}.

The relation between the geometry and morphology of deposition structures
and their formation mechanism has important applications. In recent years, technological 
advancement and the consequent access to small yet powerful computers have 
contributed greatly to simulation and numerical studies of deposition structures. 
Such studies help to develop better understanding
and control over the formation of specific forms and surfaces
suited to specific purposes. It is of practical relevance
in the fabrication of nanomaterials with important applications to
industry and medicine, such as, the manufacture of                  
sophisticated optical and electronic nanostructures and nanodevices,
magnetic carbon nanostructures for drug delivery \cite{med1} and 
smart nanostructures for monitoring, diagnoses, repair and 
treatment of human biological systems \cite{med2}.

Ballistic deposition (BD) is a simple growth model, originally proposed for describing sedimentation and aggregation in colloids \cite{vold,sutherland,famvic85,family86}. This model and its variants give rise to complex porous structures useful for studying
formation of sedimentary rock structures and dust agglomerates.

The growing surface is quantitatively expressed in terms of a 
surface width $W$, associated with the roughness of the surface,
and is defined as,
\beq
W(L,t) = \sqrt{ \frac{1}{L} \sum_{i=1}^{L} \left[ h(i,t) -
\langle h(t) \rangle \right]^2 }
\label{surfwid}
\eeq
where $L$ is the system size and $t$ is the growth time.
The surface width obeys a dynamic scaling law \cite{barabasi},
\beq
W(L,t) \sim L^{\alpha}f\left(\frac{t}{L^{z}}\right)
\label{scalinglaw}
\eeq
The exponents $\alpha$ and $\beta$ describe the growth of surface width
with system size and time.

\begin{figure}[ht]
\centering
\subfloat[]
{\includegraphics[width = 0.15\textwidth]
{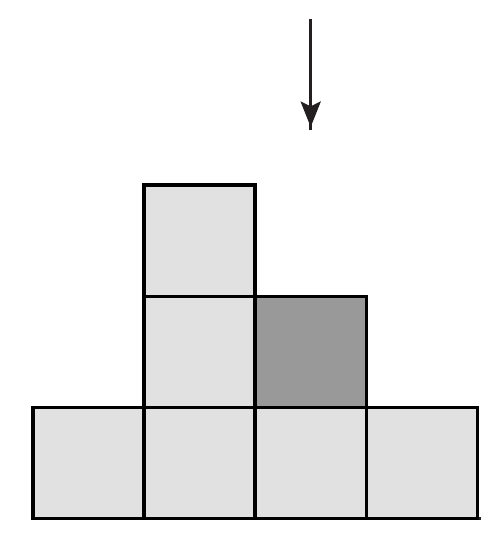}\label{fig1a}}
\subfloat[]
{\includegraphics[width = 0.11\textwidth]
{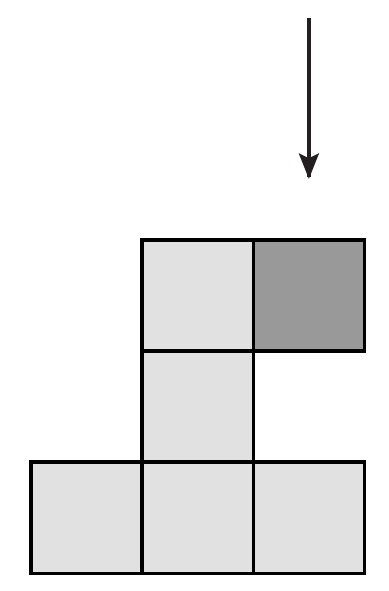}\label{fig1b}}
\subfloat[]
{\includegraphics[width=0.11\textwidth]
{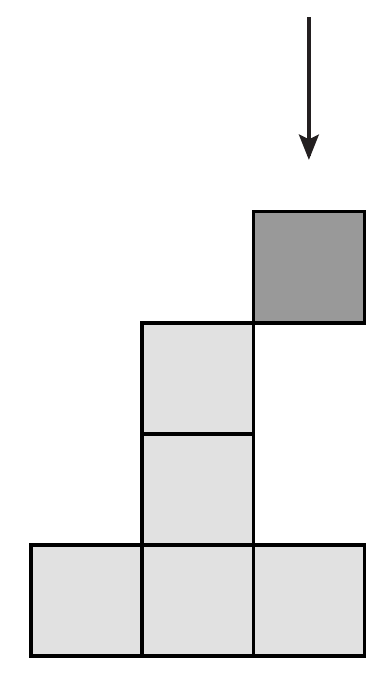}\label{fig1c}}
\caption{(a) Random deposition, (b) nearest neighbor (NN) ballistic deposition, (c) next-nearest neighbor (NNN) ballistic deposition}
\label{nnbd}
\end{figure}

In random deposition (RD), a site on the surface is selected at random.
A particle drops vertically and deposits on top of the selected column (Fig. \ref{fig1a}).
In nearest neighbor (NN) ballistic deposition, a particle 
travels in a vertical trajectory towards the randomly chosen site and
deposits onto the first surface it encounters. This may be the top of the 
chosen column or the side of the nearest neighbor column, whichever is
higher. In the next-nearest neighbor (NNN) variant of BD, the new particle deposits at the first corner or side or top of column encountered by the particle along its vertical path of descent. Fig. \ref{fig1b} and Fig. 
\ref{fig1c} illustrate the NN and NNN ballistic deposition.
In ballistic deposition, the particles are assumed to be strongly
adhering and stick to the first point of contact whereas in random deposition,
the particle deposits when it cannot go down any further.
The roughness of the growing surface grows without bound in RD, while in
BD, correlation among neighboring columns, causes the surface roughness 
to eventually saturate.
Unlike random deposition, where the surface
growth is only along the upward vertical direction, in the ballistic 
deposition, the surface grows laterally (for NN) and diagonally (for NNN)
as well.

In this work, we investigate the surface properties and bulk
structure of a modified BD model where the deposition method
interpolates between NN and NNN ballistic and the random deposition (RD)
models. The depositing particles are allowed to have varying degrees of stickiness ranging uniformly from
rigid, non-sticky to strongly adhering. The role of two types of attractive forces between the adhering particles
in the formation of the aggregate is also studied. Depending on the stickiness, 
the attractive force and the surface profile of the deposit, the incoming 
particles may stick to the corner or side
of the nearest neighbor columns or slide down to deposit on top
of the column at the randomly selected site.
It is a more realistic model for study of porous deposits formed in nature. 
\begin{figure}[ht]
\centering
\includegraphics[width=0.45\textwidth]{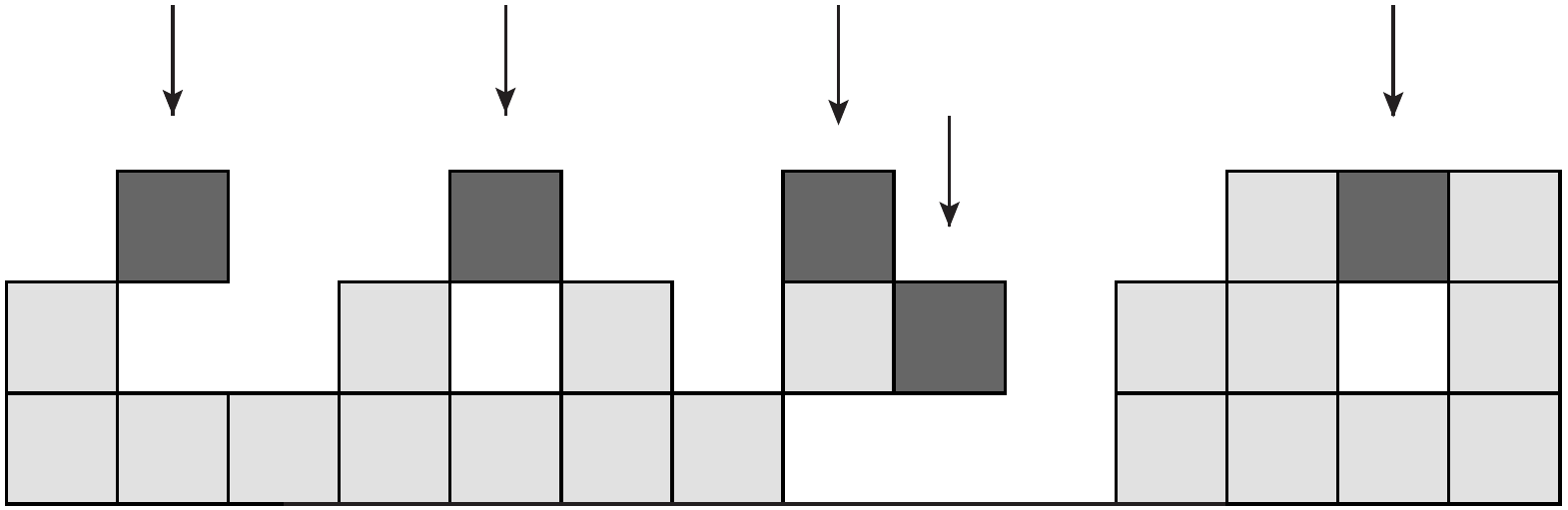}
\caption{Possible sticking positions in the present model.}
\label{deprule}
\end{figure}
The growth of surface roughness with time exhibits
four distinct regions. The stickiness of the particles is observed to have
the most dominant effect on the surface roughness, especially in the intermediate stages
of growth of deposit. The surface roughness increases at a far steeper rate than random
or KPZ \cite{KPZ86} growth. Beyond a certain crossover time, which also varies with
the stickiness parameter, the growth of surface roughness slows down to KPZ-like \cite{KPZ86}
and eventually saturates. Changing the nature of the attractive force between particles from Coulomb
to van der Waals, has negligible effect on the behavior of surface width. However, the porosity 
of the deposit is strongly influenced by the attractive force between particles as well 
as the stickiness parameter. 

\section{Formulation of the model}
\label{mbd}
In the model discussed in this article, deposition of particles takes place on a one 
dimensional substrate. Particles drop one by one vertically onto sites selected at random. Two 
factors contribute to final deposition of the particle. One is the stickiness of the particle which 
is varied by means of a parameter denoted by $a_{p}$ \cite{guptaetal}. The second  
is an inter-particle attractive force proportional to the center to center distance (denoted here by $r$)
between particles. The parameter $a_{p}$ is assigned values between $0$ and $1$, attributing stickiness to the particles. The former value corresponds to no adhesiveness and the latter to maximum stickiness. We have assumed two types of  attractive forces, the Coulomb type with an inverse square dependence on $r$ and the van der Waals type proportional to $r^{-6}$. The final sticking position of a newly 
arriving particle is decided by a sticking probability defined as 

\beq
p_{st} = \frac{a_p}{r^n } \; \; 
\label{stickrule}
\eeq
where, $n$ is $2$ for Coulomb interaction and $6$ for van der Waals interaction.
We assume the particles to be unit squares, so that the center-to-center 
distance is 1 along the side and $\sqrt{2}$ along the diagonal. A very sticky particle may 
stick to either the first encountered corner or surface of a neighboring occupied site.
A particle which is less sticky, may not stick at the first corner or surface and may slide further down, before final deposition. A new particle may deposit at the top, side or corner of an existing column depending on the sticking probability as illustrated in Fig. \ref{deprule}. Thus, in addition to vertical and lateral 
growth, the surfaces also grows along an incline of angle $45^\circ$ or $135^\circ$ to the horizontal \cite{guptaetal}. 

Simulations are performed starting with an empty substrate for system sizes $L$ $\simeq$ $32,64,128,256,512 and 1024$. A value for the sticking parameter is chosen and for a given type of inter-particle attractive force, the probabilities for corner and side sticking are determined from Eq. \ref{stickrule}. Particles are dropped onto randomly chosen sites on the substrate. If the
particle falls on a column that is higher than its nearest neighbors, it deposits onto the top of that column.
If the neighboring columns are higher, the particle may stick to the corner or side of the tallest neighboring column, provided, the corresponding sticking probability is larger than a random number generated from a 
uniform distribution between 0 and 1. Else, it slides down vertically till it encounters the next corner or surface, where a similar comparison is made. The process is repeated until deposition occurs. 

\section{Results and Discussion}
\begin{figure}[ht]
\centering
\includegraphics[width=0.45\textwidth]{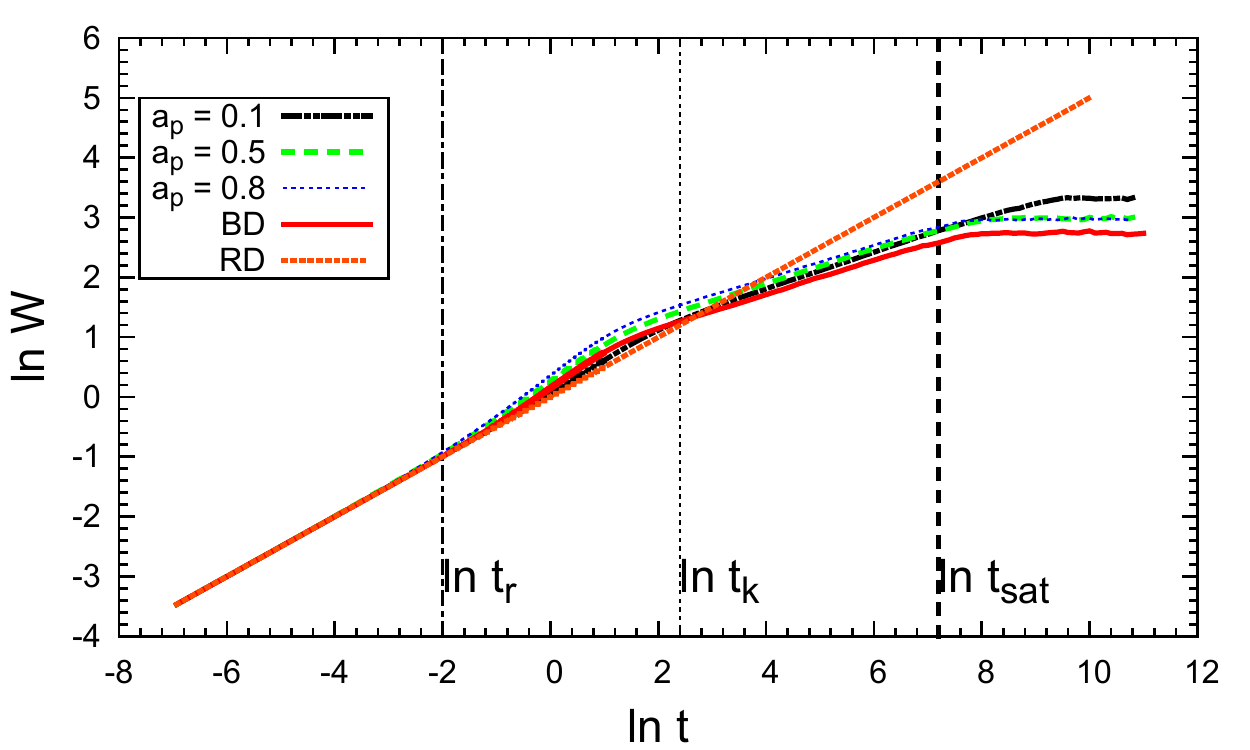}
\caption{Logarithmic plot of interface width with time for different $a_{p}$.}
\label{lnwt_ap}
\end{figure}
The numerical simulations were performed for system sizes 
mentioned above at various values of stickiness parameter ranging from 0.1 to 1.0 in 
$(1+1)$ dimension. A logarithmic plot of the interface width versus time 
at different values of $a_{p}$ is shown in Fig. \ref{lnwt_ap}.
For comparison, the plot for ballistic deposition is also shown in the 
same figure. 

At $a_p=0.0$, the particles deposit onto top of columns randomly. The surface roughness increases 
without bound as more particles deposit. 
For $a_p >0$, four distinct regions 
may be identified in the plot. An initial linear region that coincides 
with RD followed by a steep increment of interface width up to a 
certain time $t_{k}$ (see Fig. \ref{lnwt_ap}).
Thereafter, $\ln W$ increases, albeit at a much slower rate 
and eventually, beyond a time $t_{sat}$, it saturates. 
\begin{figure}[ht]
\centering
\includegraphics[width=0.45\textwidth]{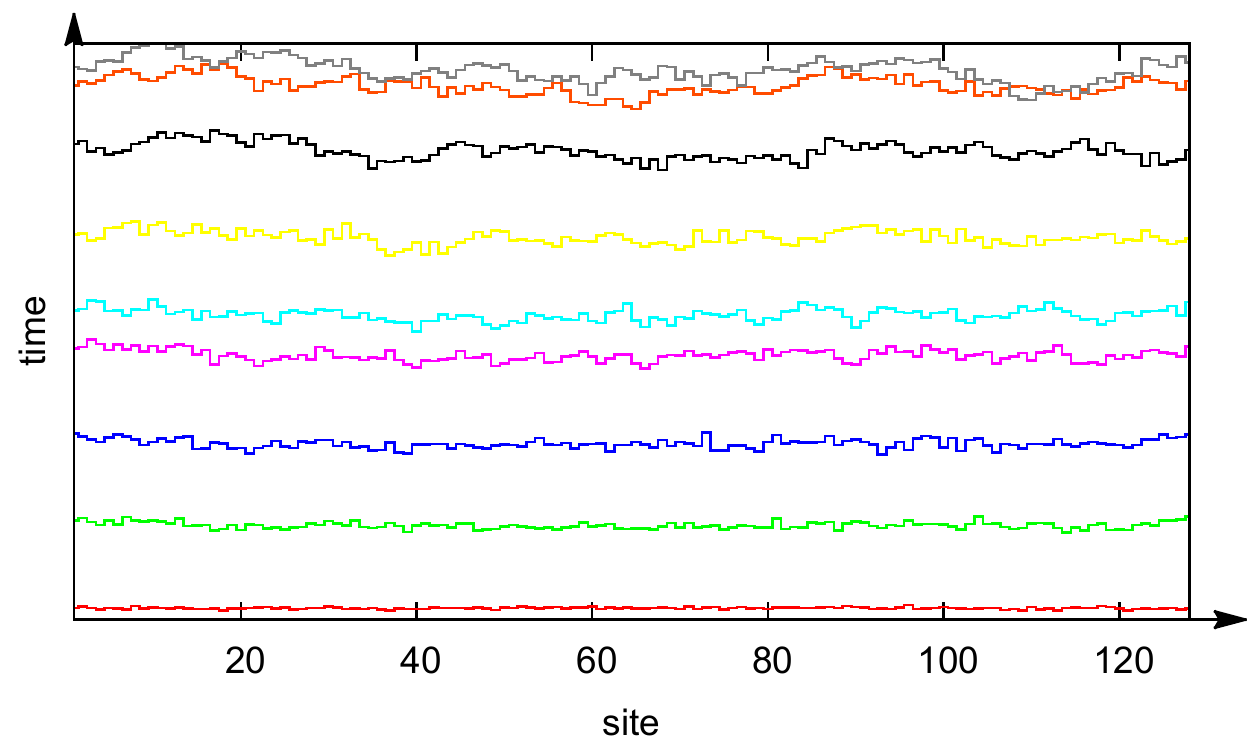} 
\caption{Evolution of the surface height in the present model}
\label{surfprof}
\end{figure}
The physical reason for this behavior 
may be understood as follows. Initially, the deposition 
process is random as there are few particles and hence for most of
the selected sites, there are no neighbors. As more particles get deposited,
new arriving particles encounter the corners and 
sides of those already deposited. The possibility 
of corner sticking results in the rapid growth of the interface width. The surface roughness thus increases with time as shown in Fig. \ref{surfprof}. 
The deviation of the interface width from random deposition is shown in Fig. \ref{eddevfromrd} for two different values of $a_{p}$. Further deposition slows down the rate of increase of surface roughness. 
\begin{figure}[ht]
\centering
\includegraphics[width=0.45\textwidth]{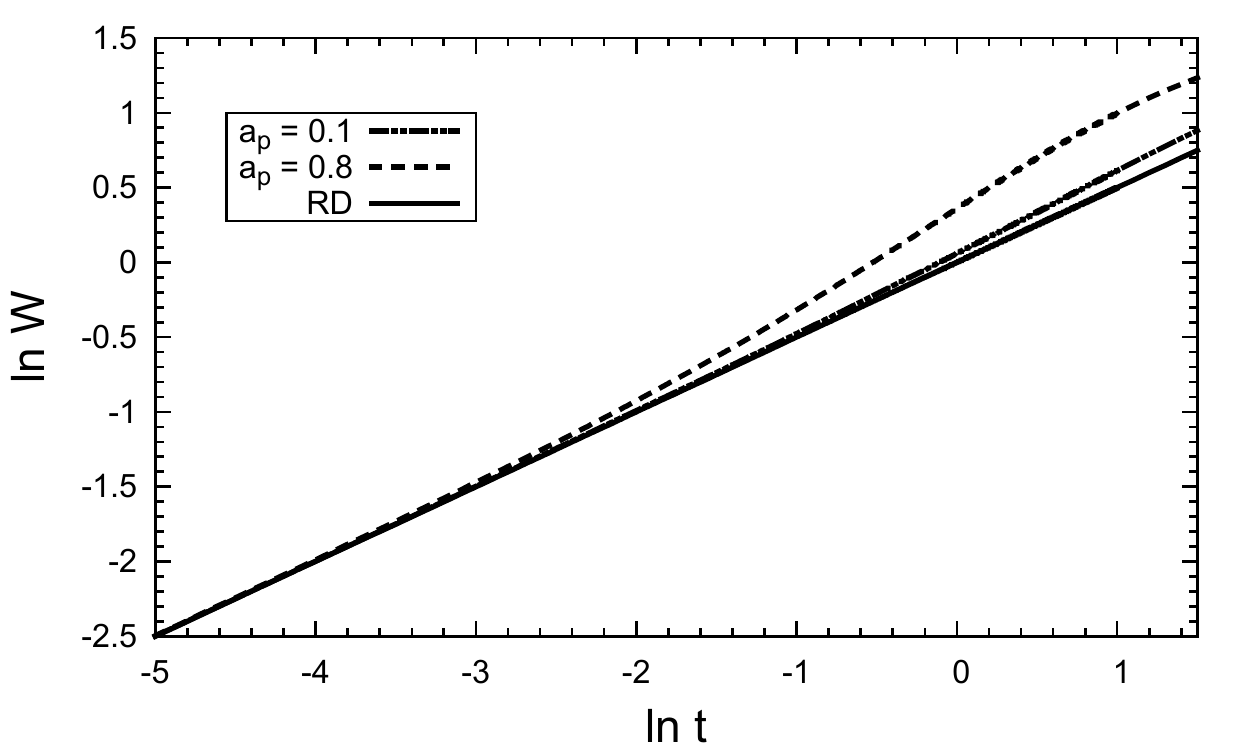}
\caption{Deviation from random growth for early times for $a_p=0.5$ and $a_p=0.8$.}
\label{eddevfromrd}
\end{figure}
This deviation of the interface width increases with $a_{p}$ as the possibility of corner sticking increases. 
For example, when $a_{p}$ = $0.1$, the probability for corner sticking 
is $0.05$ and that for side sticking is $0.1$, whereas at $a_{p}$ = $0.5$, 
the corner sticking probability increases to 0.25 and that for 
side sticking increases to 0.5. Chances of sticking to corner or side, both increase with $a_{p}$. 
Hence, the growth along both lateral and diagonal directions increases 
with $a_{p}$. 
\begin{figure}[ht]
\centering 
\includegraphics[width=0.45\textwidth]{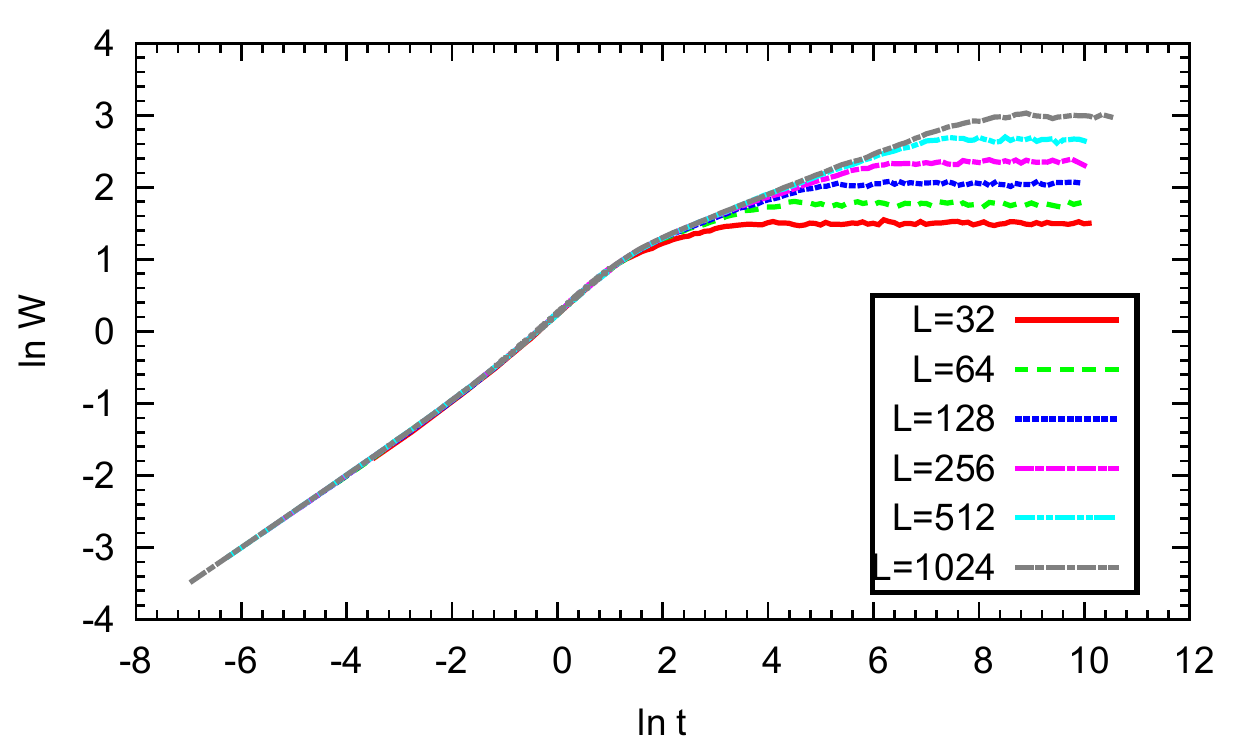} 
\caption{Variation of interface width with time for different system sizes
with $a_{p}=0.5$.}
\label{lnwap_fixed}
\end{figure}

\begin{figure}[ht]
\centering 
\includegraphics[width=0.45\textwidth]{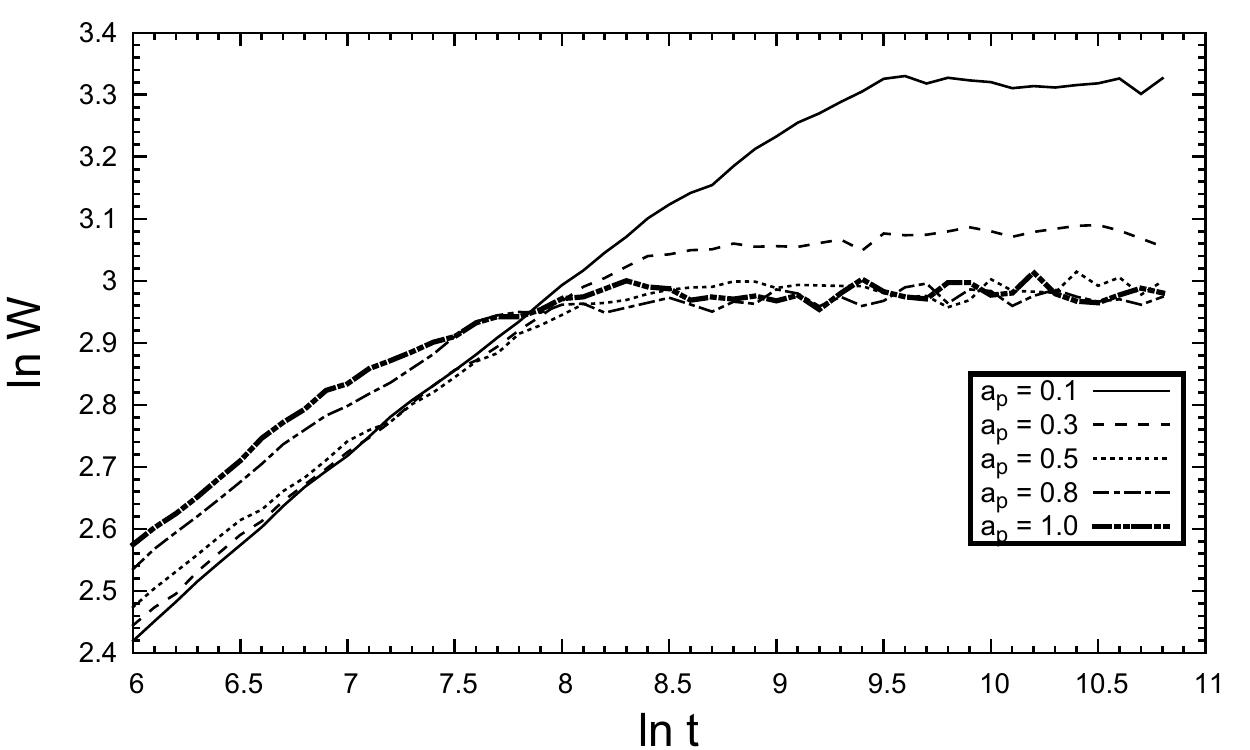} 
\caption{Variation of interface width with time in the saturated region for $L=1024$.}
\label{satreg_1024}
\end{figure}

With further deposition, the increase in surface roughness slows down. Unlike the previous region, this second slower growth region shows a decreasing slope with increase in $a_{p}$. Increase in $a_{p}$, causes more correlations among neighboring columns as
the chances of sticking to corner(s) as well as to side(s) increases.
As a result, the interface grows at a slower rate.
\begin{figure}[ht]
\centering
\includegraphics[width=0.45\textwidth]{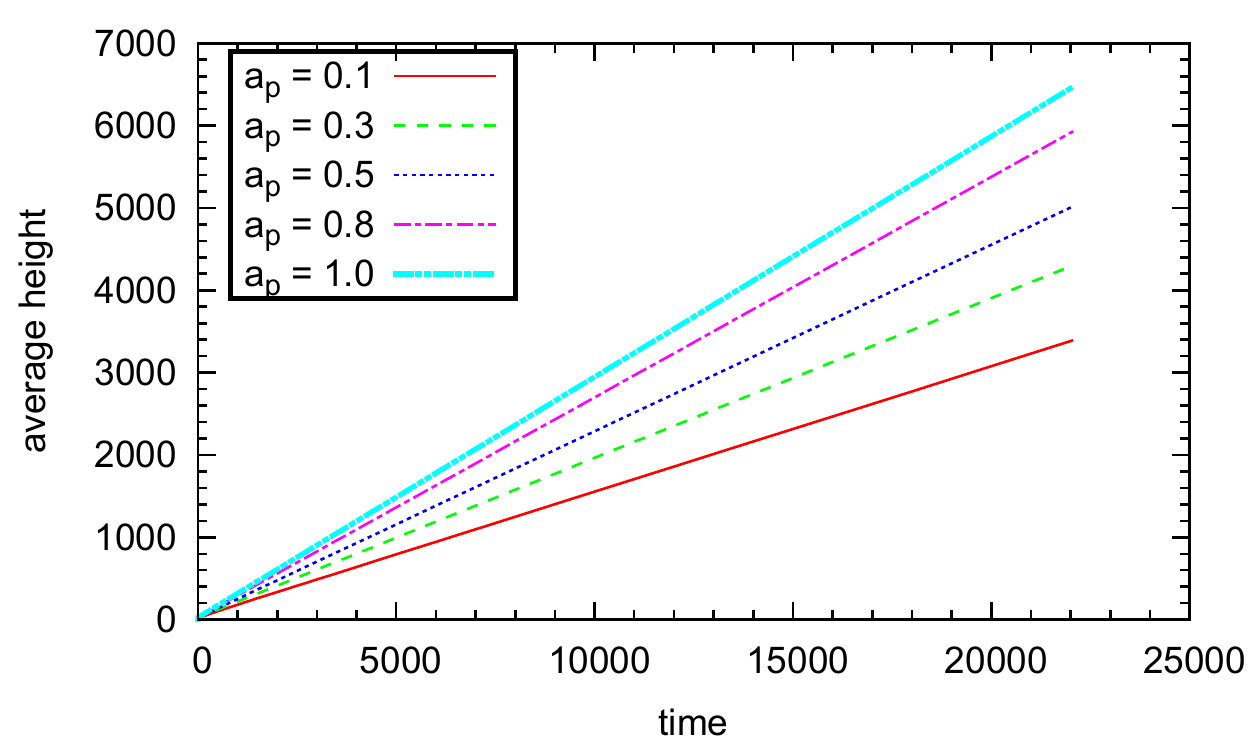}  
\caption{Evolution of average height at different $a_{p}$ for $L = 256$ and Coulomb type interaction.}
\label{avgh}
\end{figure}

\begin{figure}[ht]
\centering
\includegraphics[width=0.45\textwidth]{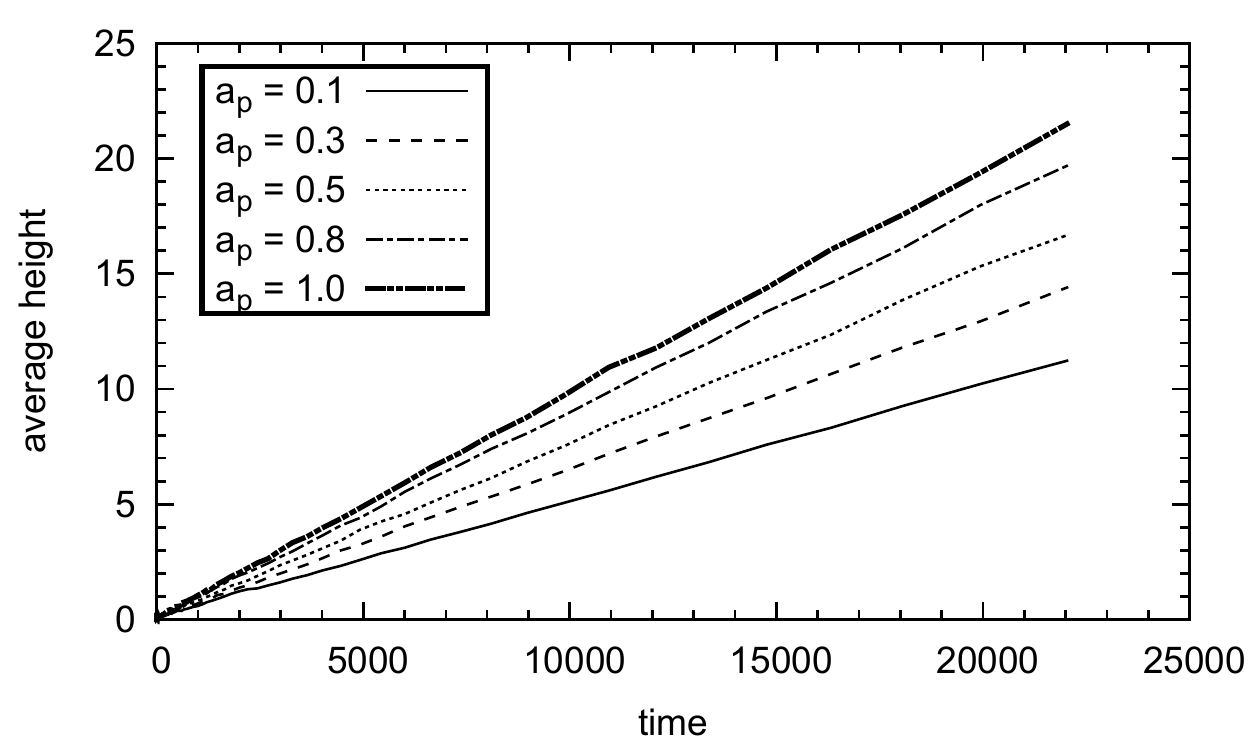}  
\caption{Evolution of average height at different $a_{p}$ for $L = 256$ and van der Waals type interaction.}
\label{avghvand}
\end{figure}
 
Thereafter, with further deposition of particles, the interface width
saturates after a time $t_{sat}$.
The saturated value of the interface width varies with both system size 
and $a_{p}$. 
For a given value of $a_{p}$, the saturated width W$_{sat}$
and the time of saturation, $t_{sat}$ increase with system size. For a fixed $a_{p}$, the evolution of the interface width
for different system sizes is shown in Fig. \ref{lnwap_fixed}.

\begin{figure}[ht]
\centering
{\includegraphics[width=0.45\textwidth]{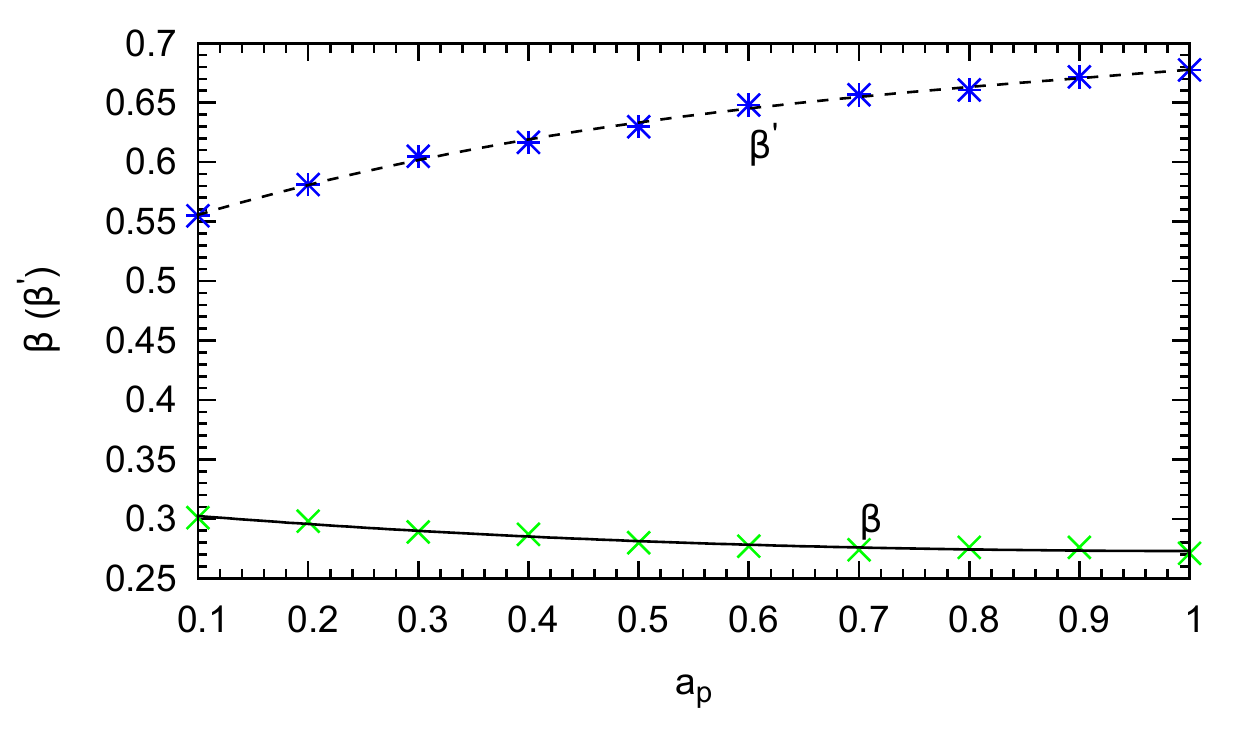}}
\caption{Variation of $\beta$ (solid line) and $\beta^{'}$ (dashed line) with $a_{p}$ 
for $L = 1024$}
\label{betaprimeap}
\end{figure}

\begin{figure}[ht]
\centering
{\includegraphics[width=0.45\textwidth]{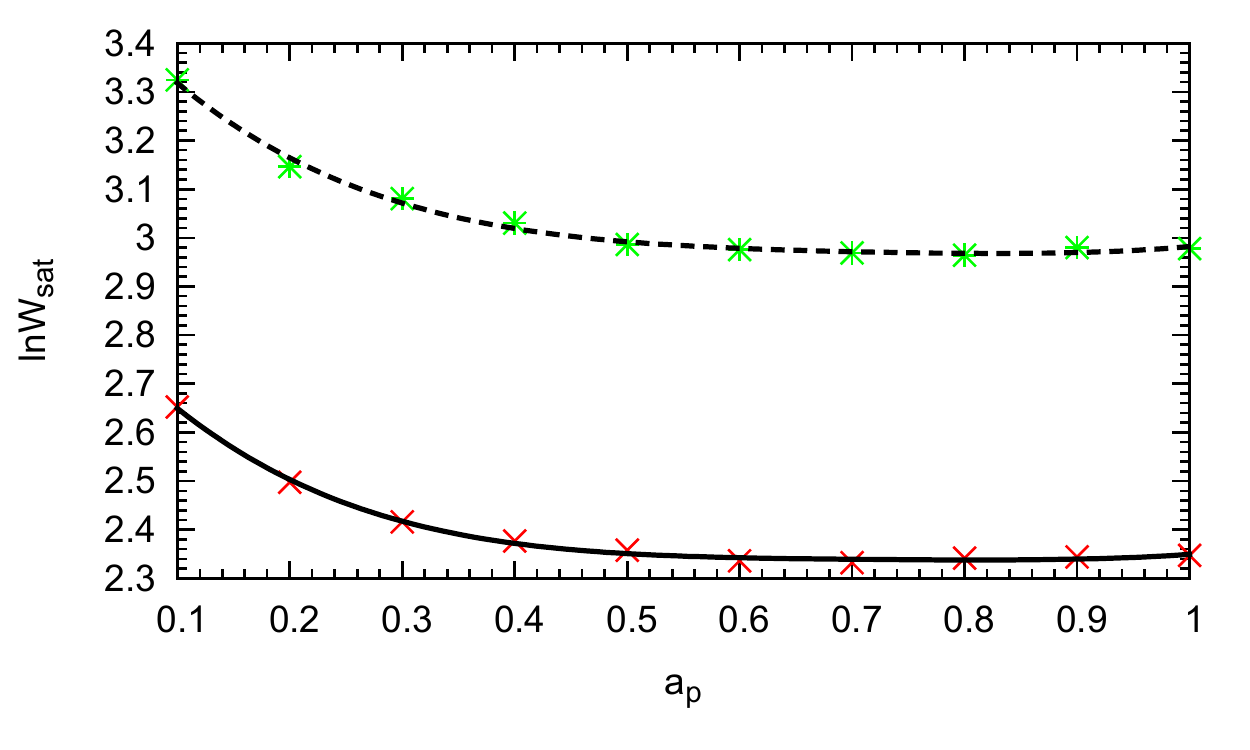}}  
\caption{Saturated width versus $a_{p}$ for $L = 256$ (solid line) and $1024$ (dashed line).}
\label{wsatap}
\end{figure}

\begin{figure}[ht]
\centering
\includegraphics[width=0.45\textwidth]{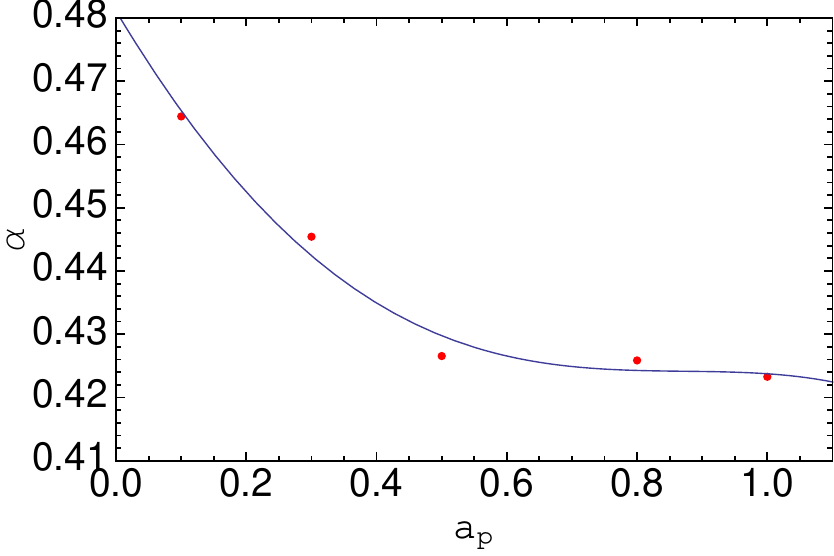}
\caption{Variation of $\alpha$ with $a_{p}$}
\label{alpap}
\end{figure}

With the system size kept fixed, the saturated width decreases with increase 
in $a_{p}$ and so does the time of saturation. There is a marked decrease in saturated width when $a_{p}$ is increased from $0.1$ to $0.3$. Thereafter, the value of the saturated width shows a very small decrease with further increase in $a_{p}$. This behavior is due to the increase in correlation length as the stickiness parameter increases. However, the correlation length cannot exceed the system size.
Hence the decrease in saturated width is most prominent at low values of $a_p$.
\begin{figure}[ht]
\centering
\includegraphics[width=0.45\textwidth]{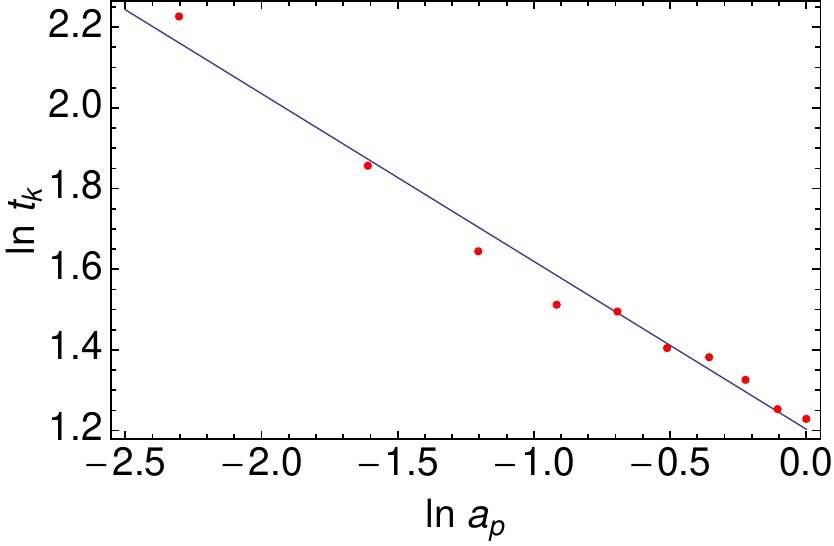}
\caption{Logarithmic plot of crossover time $t_k$ with $a_{p}$.}
\label{tkap}
\end{figure}
Fig. \ref{satreg_1024} shows the variation of the saturated interface width $a_{p}$ for system size $1024$ and Coulomb type interaction. It is also observed that the interface width saturates at earlier time with increase in $a_{p}$. When the inter-particle force is switched to van der Waals type, there is negligible change in growth and the saturation of interface width.
Fig. \ref{avgh} and \ref{avghvand} show the evolution of the average height of the surface for Coulomb and van der Waals type interaction respectively. Though the qualitative nature of the above plots are the same, the average surface height for van der Waals type of interaction is much less  than that for Coulomb type interaction.
However, for both type of forces, the average height at any time is larger and grows faster
for larger values of $a_p$. 

\begin{table}[htbp]
\caption{Values of $\beta^{'}$ for different $a_{p}$}
\label{tab2}
\begin{center}
\begin{tabular}{cccccc}
\hline \hline
$L$ & $a_{p} = 0.1$ & $a_{p} = 0.3$ & $a_{p} = 0.5$ & $a_{p} = 0.8$ & $a_{p} = 1.0$ \\
\hline 
128 & 0.55 & 0.60 & 0.63 & 0.66 & 0.67 \\
256 & 0.55 & 0.60 & 0.63 & 0.67 & 0.67 \\
512 & 0.55 & 0.60 & 0.63 & 0.66 & 0.68 \\
1024 & 0.55 & 0.60 & 0.63 & 0.66 & 0.68 \\
\hline \hline
\end{tabular}
\end{center}
\end{table}

\begin{figure}[ht]
\centering
\includegraphics[width=0.45\textwidth]{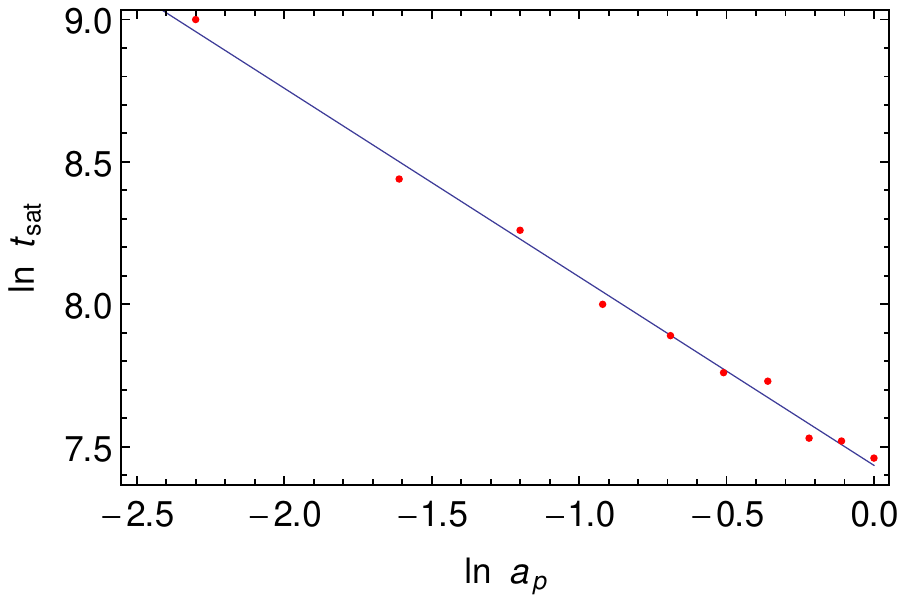}
\caption{Saturation time $t_{sat}$ versus $a_{p}$ 
in log-log scale.}
\label{tsatap}
\end{figure}

For $t \ll 1$, the roughness of interface grows with exponent $\sim 0.5$ as for random growth. 
Further deposition causes a sharp increase in the roughness. The 
values of this growth exponent, denoted by $\beta'$, 
are tabulated below (Table \ref{tab2}) for different 
system sizes and $a_{p}$. For a fixed system size, $\beta^{'}$
increases with $a_p$. However, at any given value of $a_p$, $\beta^{'}$
does not change with the system size. This behavior is observed for both Coulomb type and van der Waals type attractive forces.

\begin{figure}[ht]
\centering
\includegraphics[width=0.45\textwidth]{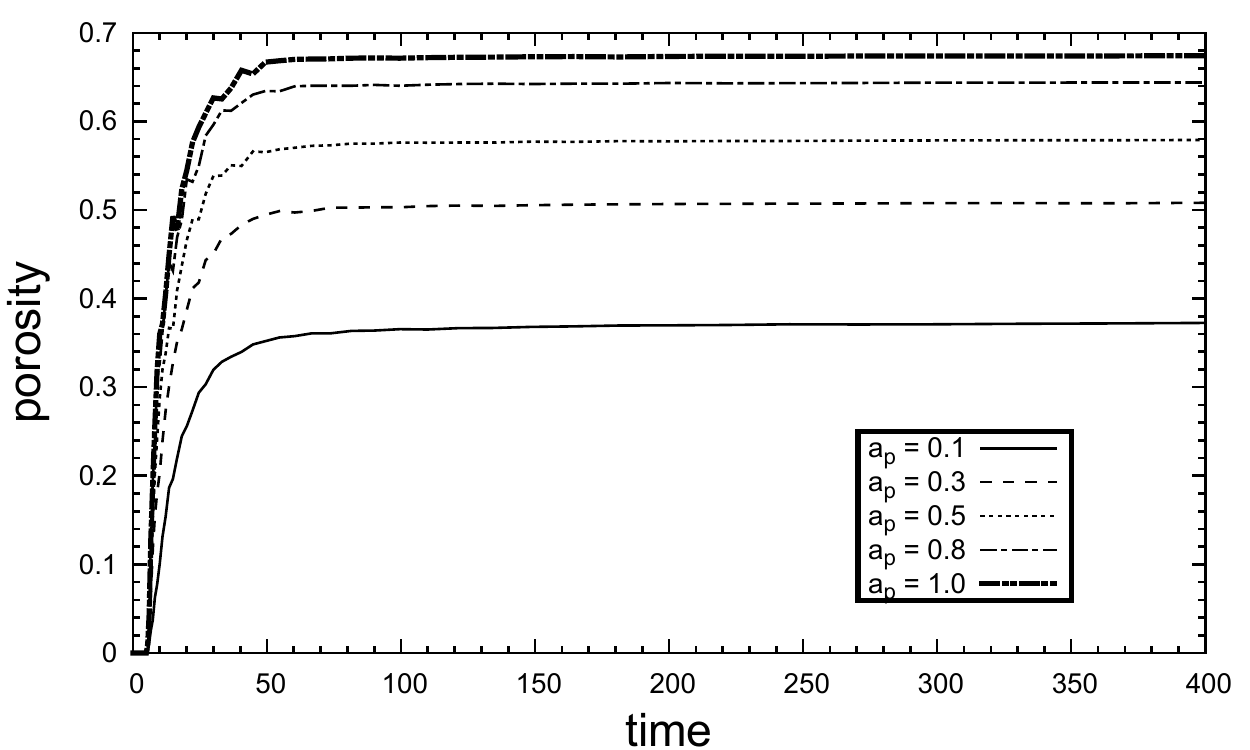}  
\caption{Growth of porosity at different $a_{p}$}
\label{pore}
\end{figure}

This rapid growth of roughness is not sustained beyond a certain time $t_{k}$. The rate of increase of roughness slows down to a KPZ like growth with a different exponent denoted by $\beta$. 
This exponent, unlike $\beta'$, decreases with $a_p$. For a 
fixed $a_p$, $\beta$ increases with the system size approaching a value $0.31$.
In Table \ref{tabalbe}, values of $\beta$ (exponent for KPZ like growth) are tabulated for different system sizes and $a_{p}$.

\begin{figure}[ht]
\centering
\includegraphics[width=0.45\textwidth]{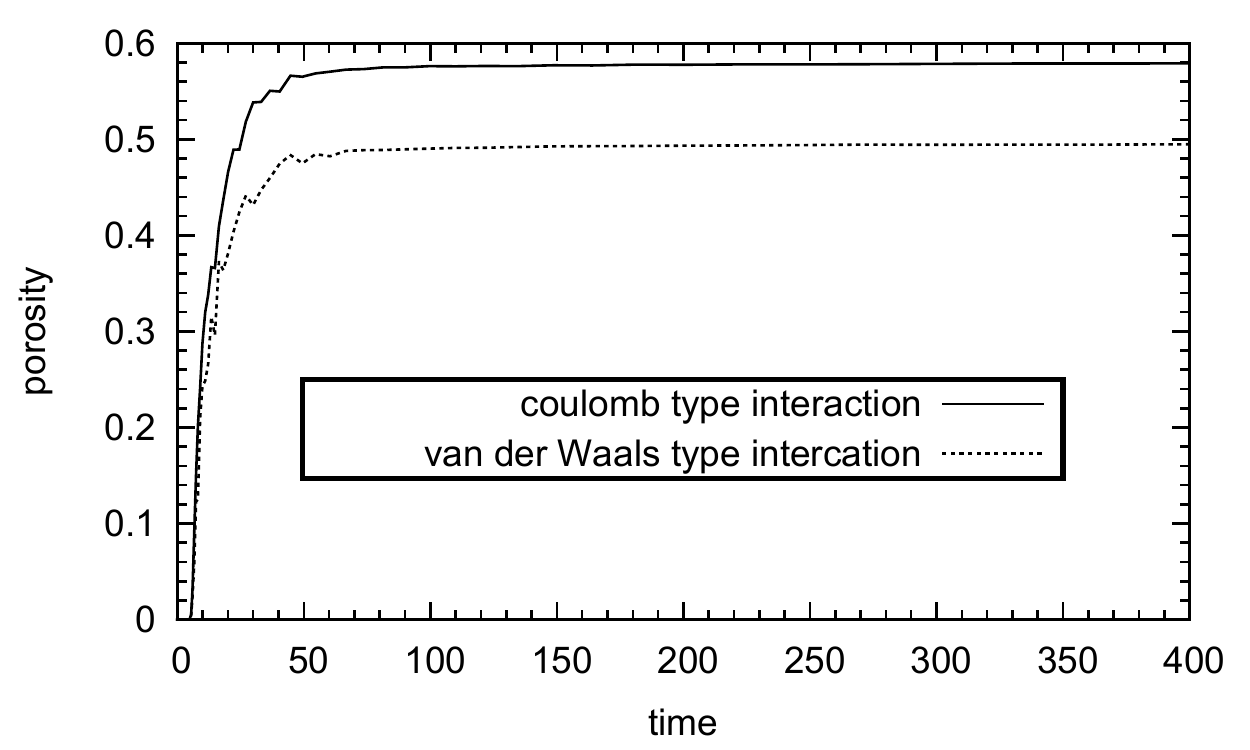}  
\caption{Growth of porosity for Coulomb and van der Waals type interaction.}
\label{comppore}
\end{figure}

In Fig. \ref{betaprimeap}, $\beta$ and $\beta'$ are plotted versus $a_p$ for system size 1024 and a third degree polynomial.

\begin{figure}[ht]
\centering
\includegraphics[width=0.45\textwidth]{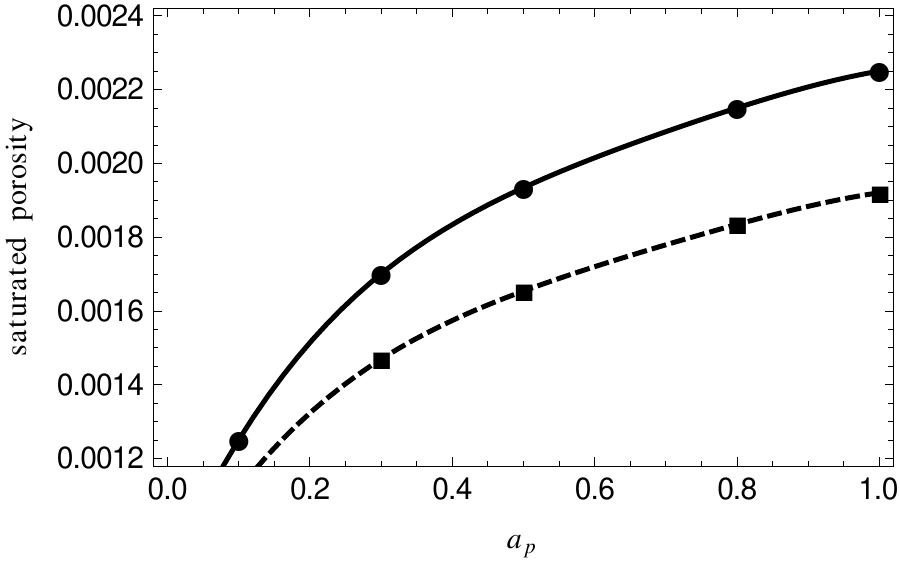}  
\caption{Saturated Porosity variation with $a_{p}$ for Coulomb (solid line) and van der Waals type (dashed line) interaction.}
\label{satpore}
\end{figure}

The saturated width at any value of $a_{p}$ in the present model is larger than the saturated width in ballistic deposition. At any value of $a_p$, saturated width is larger for larger system sizes as shown in Fig. \ref{wsatap}. The saturated width W$_{sat}$ approaches a limiting value as $a_p \to 1$ for a fixed system size.
This limiting value is size dependent as shown in Fig. \ref{wsatap}.
 
\begin{table}[htbp]
\caption{Values of $\beta$ for different $a_{p}$}
\label{tabalbe}
\begin{center}
\begin{tabular}{cccccc}
\hline \hline
$L$ & $a_{p} = 0.1$ & $a_{p} = 0.3$ & $a_{p} = 0.5$ & $a_{p} = 0.8$ & 
$a_{p} = 1.0$ \\
\hline 
128 & 0.2911 & 0.271721 & 0.250261 & 0.24836 & 0.228566 \\
256 & 0.289875 & 0.286438 & 0.261282 & 0.25338 & 0.247328 \\
512 & 0.29726 & 0.287251 & 0.270889 & 0.265053 & 0.259155 \\
1024 & 0.298774 & 0.297451 & 0.280052 & 0.275126 & 0.270122 \\
\hline \hline
\end{tabular}
\end{center}
\end{table}

Fig. \ref{alpap} reveals a power law variation of the roughness exponent
$\alpha$ with $a_p$. The variation appears to follow a fourth order polynomial.

Corresponding to the four distinct regions, three transition times may be
associated. The transition to the rapid growth region from the initial
linear RD region is denoted by $t_r$. 
This first transition time, for a given system, is observed to decrease 
with increase in $a_{p}$. The deviation from RD is due to the fact that the value of the variance determining the roughness of the interface width increases as particles stick more to corners.  
Increase in $a_p$ translates into increased possibility of sticking to corners.
Hence, the deviation from RD begins sooner. 
With further deposition of particles, the growth of interface width slows down
and shows a KPZ-like growth. The time for this transition is denoted by $t_k$.

Fig. \ref{tkap} shows the power law dependence of crossover time 
$t_{k}$ on $a_{p}$. The logarithmic plot shows $t_{k} \sim a_{p}^{-0.4\pm0.04}$ for large system size. 
The saturated crossover time, $t_{sat}$, scales both with the system size $L$ and $a_{p}$ (\ref{tsatap}). It is observed that,
\begin{equation}
t_{sat} \sim L^{z} \;\;\; {\rm where} \;\;\; z = \frac{\alpha}{\beta}
\;\;\; {\rm and} \;\;\; t_{sat} \sim a_{p}^{0.7 \pm 0.03}
\end{equation} 
The $z$ value obtained by the linear fit of $t_{sat}$ with system size is 
$\sim$ 1.4.

Vacancies or holes in the bulk of the deposit give rise to a porous
structure. The porosity may be defined as the fraction of holes (unoccupied
sites) in the deposit. 
For Coulomb type interaction between particles, it is observed that, for any given value of $a_{p}$, the porosity 
rapidly increases with time and then saturates as shown in Fig. \ref{pore}. For van der Waals type of interaction, the qualitative nature of the plot is the same. Quantitatively however, the value of porosity is less than that for Coulomb type interaction as shown in Fig. \ref{comppore}.
The saturated porosity is found to be independent of the system size. For a fixed system size, it increases with $a_{p}$. The variation of saturated porosity with $a_{p}$ for both Coulomb and van dr Waals type of interaction is depicted in Fig. \ref{satpore}. The deposit structure formed with van der Waals type of interaction is less porous. A fourth order polynomial gives a good fit to the plot of saturated porosity.

\section{Conclusion}

To summarize, we have studied the bulk and surface properties of deposits formed by particles with varying degrees of stickiness and different inter-particle attractive forces. The stickiness parameter has a more dominant effect on the surface roughness whereas the porosity is rather strongly influenced by the nature of the inter-particle attractive force. In the intermediate stages of growth the surface roughness shows a far steeper rate of increase than in the random or KPZ growth. It would be interesting to derive a continuum stochastic equation corresponding to this growth model and ascertain the role of sticking probability on the coefficient of the various terms of the stochastic equations.


\end{document}